\documentclass[secnumarabic, graphics,floatfix, nofootinbib,tightenlines,nobibnotes, aps, prl, 12pt]{revtex4-1}
\usepackage{graphicx}
\usepackage[english]{babel}
\usepackage[utf8]{inputenc}
\usepackage{amsmath,amssymb}
\usepackage{amsfonts}
\usepackage{multirow}
\usepackage{pstricks}
\usepackage{float}
\usepackage{subfigure}
\usepackage{color}
\usepackage{epsfig}
\usepackage{pst-tools}
\usepackage{mnsymbol}

\begin{document}

\title{Thermodynamical consistency of quasiparticle model at finite baryon density}

\author{Hong-Hao Ma$^1$}
\author{Kai Lin$^{2,3}$}
\author{Wei-Liang Qian$^{3,1,4}$}
\author{Yogiro Hama$^{5}$}
\author{Takeshi Kodama$^{6,7}$}

\affiliation{$^1$ Faculdade de Engenharia de Guaratinguet\'a, Universidade Estadual Paulista, 12516-410, Guaratinguet\'a, SP, Brazil}
\affiliation{$^2$ Institute of Geophysics and Geoinformatics, China University of Geosciences, 430074, Wuhan, Hubei, China}
\affiliation{$^3$ Escola de Engenharia de Lorena, Universidade de S\~ao Paulo, 12602-810, Lorena, SP, Brazil}
\affiliation{$^4$ Center for Gravitation and Cosmology, College of Physical Science and Technology, Yangzhou University, Yangzhou 225009, China}
\affiliation{$^5$ Instituto de F\'isica, Universidade de S\~ao Paulo, C.P. 66318, 05315-970, S\~ao Paulo-SP, Brazil}
\affiliation{$^6$ Instituto de F\'isica, Universidade Federal do Rio de Janeiro, C.P. 68528, 21945-970, Rio de Janeiro-RJ , Brazil}
\affiliation{$^7$ Instituto de F\'isica, Universidade Federal Fluminense, 24210-346, Niter\'oi-RJ, Brazil}

\date{July 3, 2019}

\begin{abstract}

In this work, we revisit the thermodynamical self-consistency of quasiparticle model with finite baryon chemical potential adjusted to lattice QCD calculations.
Here, we investigate the possibility where the effective quasiparticle mass is also a function of its momentum, $k$, in addition to temperature $T$ and chemical potential $\mu$.
It is found that the thermodynamic consistency can be expressed in terms of an integro-differential equation concerning $k$, $T$, and $\mu$.
We further discuss two special solutions, both can be viewed as sufficient condition for the thermodynamical consistency, while expressed in terms of a particle differential equation.
The first case is shown to be equivalent to those previously discussed by Peshier {\it et al}.
The second one, obtained through an {\it ad hoc} assumption, is an intrinsically different solution where the particle mass is momentum dependent.
These equations can be solved by using boundary condition determined by the lattice QCD data at vanishing baryon chemical potential.
By numerical calculations, we show that both solutions can reasonably reproduce the recent lattice QCD results of the Wuppertal-Budapest and HotQCD Collaborations, and in particular, those concerning finite baryon density.
Possible implications are discussed.

\pacs{12.38.Bx, 12.38.Aw, 11.15.Bt}

\end{abstract}

\maketitle

\section{I Introduction}

Quasiparticle approach is part of the efforts to understand the physics of the quark-hadron transition characterized by a dramatic change in the number of degrees of freedom where nonperturbative effects are dominant.
The model provides a reasonable, as well as phenomenological, description of the thermodynamic properties of quark-gluon plasma (QGP) which deviate significantly from those of an ideal gas of non-interacting quarks and gluons.
The success of the quasiparticle picture thus strengthens the notion of quasiparticle ansatz.
It may further open up new possibilities for the development of effective theories from a more fundamental viewpoint, concerning the underlying physics of QGP, which is nonperturbative in nature.
Indeed, as indicated by lattice quantum chromodynamics (QCD) calculations, the QGP pressure and energy density deviate by about 15-20\% from the Stefan-Boltzmann limit even at temperatures $T > 3 T_c$~\cite{latt-review-01}.
On the other hand, the square of the speed of sound, $c_s^2$, extracted from lattice QCD, is smaller than that of an ideal gas of massless particles.
In particular, it is found that as the temperature decreases while the system approaches the transition region, $c_{s}^{2}$ reaches down to a minimum and then increases again in accordance with the hadronic resonance gas (HRG) description of the system~\cite{qgp-review-03}.
Since these thermodynamical properties may lead to observable consequences through their impact on the hydrodynamically expanding phase during the relativistic heavy ion collisions, they are, therefore, essential features in the study of the strongly interacting QGP matter.
The lattice QCD, as an exact and yet numerical technique to obtain the equation of state (EoS), it is still challenging to study finite density QCD in large baryon density and low temperature regions.
Besides, there are other attempts to investigate thermal properties of the QGP such as dimensional reduction~\cite{qcd-phase-DR-01,qcd-phase-DR-02,qcd-phase-DR-03}, hard thermal loop (HTL) resummation scheme~\cite{qcd-HTL-01,qcd-HTL-02,qcd-HTL-03,qcd-HTL-04,qcd-HTL-05,qcd-HTL-06,qcd-HTL-07}, Polyakov-loop model~\cite{qcd-phase-PL-01,qcd-phase-PL-02}, as well as approaches in terms of hadronic degrees of freedom~\cite{qcd-phase-Sigma-01,qcd-phase-Sigma-02,qcd-phase-Sigma-03}.
The subtlety among different approaches is how to appropriately tackle the nonperturbative regime of QCD, which, in particular, as the temperature decreases and approaches $T_c$, still cannot be accurately described to date.

Inspired by its counterparts in other fields of physics, the quasiparticle ansatz assumes that the strongly interacting matter consists of non-interacting quanta which carry the same quantum numbers of quarks and gluons.
The strong interactions between the elementary degrees of freedom are incorporated through the medium dependent quasiparticle mass.
The quasiparticle approach was first introduced by Peshier $et~al.$~\cite{eos-quasiparticle-13} for the description of gluon plasma, where the temperature dependent particle mass was proposed.
However, it was subsequently pointed out by Gorenstein and Yang~\cite{eos-quasiparticle-gorenstein-01} that thermodynamic quantities evaluated by using ensemble average may not agree with those obtained by thermodynamic relations.
The issue can be resolved by reformulating the thermodynamics of the quasiparticle model through the requirement of an exact cancelation between the additional contributions from the temperature dependent particle mass and those from the bag constant.
The latter is assumed to be temperature dependent and determined by the condition of thermodynamic consistency.
Thereafter, the thermodynamical consistentancy were further explored by many other authors~\cite{eos-latt-11,eos-quasiparticle-14,eos-quasiparticle-07,eos-quasiparticle-03,eos-quasiparticle-04,eos-quasiparticle-15,eos-latt-17}.

By appropriately addressing the question of gauge invariance, the effective mass of a particle can be defined either by the pole of the effective propagator or through the Debye screen mass extracted from the excitations at small momentum.
The calculations using HTL approximation show that the gluon screen mass extracted from the dispersion relation for transverse gluons~\cite{qcd-HTL-08,qcd-HTL-09} are in accordance with the Debye mass obtained at the limit of small momentum~\cite{qcd-HTL-01,qcd-HTL-02,qcd-HTL-10}.
Therefore, in practice, the specific forms of quasiparticle mass are taken as a function of temperature, chemical potential as well as the running coupling constant that are usually inspired by the HTL results.
As a further matter, the running coupling can be replaced by an effective coupling, $G^2(T,\mu)$, which in turn is determined by a flow equation~\cite{eos-latt-11,eos-quasiparticle-16,eos-latt-12,eos-latt-16}.
The latter is a partial differential equation, and its boundary condition can be chosen as the effective coupling at $\mu=0$, adjusted to the lattice QCD data.
It is shown that the thermodynamic properties obtained from lattice calculations, especially those for nonvanishing chemical potential, are described remarkably well.

In order to guarantee the thermodynamic consistency, the following relation is to be satisfied
\begin{eqnarray}
\left.\frac{\partial \ln Q_G}{\partial m}\right|_{T,\mu} = 0 . \label{gol}
\end{eqnarray}
where $Q_G$ is the the grand partition function.
In literature, it is required subsequently~\cite{eos-quasiparticle-gorenstein-01,eos-latt-11}
\begin{eqnarray}
\frac{dB}{dm} = \left.\frac{\partial p(T,\mu,m)}{\partial m}\right|_{T,\mu} . \label{go0a}
\end{eqnarray}
Here, the bag constant $B$ is understood to be a function of the particle mass $m$ only, and its temperature (and chemical potential) dependence is inherited implicitly from that of the quasiparticle mass $m=m(T,\mu)$.
It is straightforward to show that Eq.(\ref{go0a}) indeed implies to Eq.(\ref{gol}).
However, if $B$ explicitly depends on temperature, there will be an extra contribution to the thermodynamic quantities which is not accounted for by Eq.(\ref{go0a}).
By examing the r.h.s. of Eq.(\ref{go0a}), it turns out to be an explicit function of $T$, $\mu$ and $m$.
Therefore, the requirement that the r.h.s. of Eq.(\ref{go0a}) is a function of temperature (and chemical potential) only through the quasiparticle mass furnishes a more stringent condition.
In this work, we show that the above consideration leads to an integro-differential equation, which is equivalent to the flow equation introduced in the Ref.~\cite{eos-latt-11} under certain circumstances.
Moreover, we show that there are also other possibilities which accommodate the requirement for thermodynamical consistency.

The present work is organized as follows.
In the next section, we review the question concerning thermodynamical consistency in the quasiparticle model.
An integro-differential equation for quasiparticle mass is derived.
Two special solutions are discussed, both are expressed in terms of a particle differential equation, and can be solved by the method of characteristics.
We show that the first case is precisely what was derived and investigated by Peshier {\it et al}.
The second solution, on the other hand, is an intrinsically different one where particle mass is found to be a function of momentum.
The numerical results are presented in section III.
By using the lattice QCD data at $\mu=0$ as the boundary condition, we show that both solutions can reasonably reproduce the recent lattice QCD results.
In particular, the results concerning finite baryon density are presented.
The last section is devoted to discussions and concluding remarks.

\section{II Thermodynamic consistency for quasiparticle model with temperature and chemical potential dependent mass}

In this section, the thermodynamic consistency for quasiparticle model is revisited.
Our discussions are based on the quasiparticle model proposed by Begun {\it et al}.~\cite{eos-quasiparticle-gorenstein-02}.
An interesting aspect of the approach, as pointed out by the authors, is the existence of an additional free parameter.
To be specific, it is shown that pressure, while following its traditional definition in statistical physics, is determined up to an extra free parameter.

Let us first write down the expressions for energy and particle number as they are formulated as ensamble average as follows,
\begin{eqnarray}
\langle E \rangle= \frac{\sum\limits_i E_i \exp (-\alpha N_i-\beta E_i)}{\sum\limits_i \exp (-\alpha N_i-\beta E_i)} , \nonumber\\
\langle N \rangle= \frac{\sum\limits_i N_i \exp (-\alpha N_i-\beta E_i)}{\sum\limits_i \exp (-\alpha N_i-\beta E_i)} .\label{ensembleavg}
\end{eqnarray}
where the ensamble average is carried out among all possible microscopic state $i$ of the system, and $N_i$ and $E_i$ are respectively the total number and total energy of the state in question.
The above expression can be rewritten in terms of the grand partition function,
\begin{eqnarray}
Q_{G}=\langle \exp[-\alpha \hat{N}-\beta \hat{H}_{\mathrm{eff}}]\rangle ,
\end{eqnarray}
where
\begin{eqnarray}
\hat{H}_{\mathrm{eff}}=\hat{H}_{\mathrm{id}}+E_0+E_1 .
\end{eqnarray}
Here $\hat{H}_{\mathrm{id}}$ is the Hamiltonian of ideal gas of quasiparticles
\begin{eqnarray}
\hat{H}_{\mathrm{id}}=\sum_j\sum_{\mathbf{k}}\omega(\mathbf{k})a_{\mathbf{k},j}^\dagger a_{\mathbf{k},j} , \label{ehid}
\end{eqnarray}
where $j$ corresponds to internal degrees of freedom.
Here $E_0$ is a temperature and chemical potential dependent function associated with the bag constant $B$ proposed by Gorenstein and Yang~\cite{eos-quasiparticle-gorenstein-01}.
This term is used to cancel out the effects of the temperature (and chemical potential) dependence of the quasiparticle mass through Eq.(\ref{gol}), to be discussed further below.
$E_1$ is the above-mentioned free parameter, which is proportional to the temperature.
The $E_1$ term is singled out from $E_0$ owing to its peculiar nature.
As shown below, it allows one to further adjust the value of the pressure for any given energy density~\cite{eos-quasiparticle-gorenstein-02}.

Quasiparticle ansatz assumes that one may carry out the calculations in the momentum space where the Hamiltonian is diagonal.
To be more specific, one makes the following substitutions for the ideal gas part
\begin{eqnarray}
\sum_j\sum_{\mathbf{k}} \rightarrow \frac{gV}{(2\pi)^3}\int d\mathbf{k} .
\end{eqnarray}
where $g$ is the degeneracy factor.
Now, thermodynamical quantities can also be expressed regarding the derivatives of the grand partition function.
For instance, the energy density reads
\begin{eqnarray}
\varepsilon=\frac{\langle E\rangle}{V}=-\frac{1}{V}\frac{\partial \ln Q_G}{\partial\beta}=\epsilon_{\mathrm{id}}+\frac{E_0}{V}+\frac{E_1}{V}+\frac{1}{V}\langle \beta\frac{\partial E_1}{\partial\beta}\rangle = \epsilon_{id} + B. \label{energydensity}
\end{eqnarray}
where
\begin{eqnarray}
\epsilon_{\mathrm{id}}=\frac{g}{2\pi^2}\int_0^\infty \frac{k^2dk\omega^*(k,T,\mu)}{\exp[(\omega^*(k,T,\mu)-\mu)/T]\mp 1}+\mathrm{c.t.} \,\,,
\end{eqnarray}
with on-shell dispersion relation
\begin{eqnarray}
\omega^*(k,T,\mu)=\sqrt{m(T,\mu)^2+k^2} , \label{disponshell}
\end{eqnarray}
and $B = \lim_{V\rightarrow \infty}\frac{E_0}{V} $ is the bag constant and the counter term ``$\mathrm{c.t.}$" indicates contributions from anti-particles obtained by the substitution $\mu\rightarrow -\mu$ in the foregoing term.
Here, the contribution from the temperature dependence of quasiparticle mass has already been canceled out with the temperature dependence of $E_0$.
If the system has vanishing chemical potential $\mu = 0$, one has $B=B(\mu=0, T)\equiv B(T)$ and $m=m(\mu=0, T)\equiv m(T)$, in general, one can invert the second function to find $T=T(m)$ and express $B$ as a function of $m$.
Thus the above requiement Eq.(\ref{gol}) regarding $E_0$ implies
\begin{eqnarray}
\frac{dB}{dm} = -\frac{gm}{2\pi^2}\int_0^\infty \frac{k^2dk}{\omega^*(k,T)}\frac{1}{\exp[(\omega^*(k,T))/T]\mp 1} . \label{go1}
\end{eqnarray}
At finite baryon density, however, one is dealing with a bivariate function $B=B(\mu, T)$.
Thus the above argument is not valid.
In general, $B$ may explicitly depend on $T$ besides its dependence through $m$, but one still can write down
\begin{eqnarray}
\frac{\partial B}{\partial T}= -\frac{g}{2\pi^2}\int_0^\infty \frac{k^2dk}{\omega^*(k,T,\mu)}\frac{1}{\exp[(\omega^*(k,T,\mu)-\mu)/T]\mp 1}m\frac{\partial m}{\partial T} + \mathrm{c.t.}\,\,. \label{go2}
\end{eqnarray}
Furthermore, since $E_1$ is linear in $1/\beta$, one has $\langle\beta\frac{\partial E_1}{\partial\beta}\rangle=\beta\frac{\partial E_1}{\partial\beta}=-E_1$, thus the last equality of Eq.(\ref{energydensity}) is justified.

Similarly, the pressure is interpreted as a ``general force", which reads
\begin{eqnarray}
p=\frac{1}{\beta}\frac{\partial \ln Q_G}{\partial V}=\frac{1}{\beta}\frac{\ln Q_G}{V}=p_{\mathrm{id}} - B - \frac{E_1}{V} \label{pressure},
\end{eqnarray}
where
\begin{eqnarray}
p_{\mathrm{id}}&=&\frac{\mp g}{2\pi^2}\int_0^\infty k^2dk\ln \left\{ 1\mp\exp\left[\left(\mu-\omega^*(k,T,\mu)\right)/T\right]\right\}+\mathrm{c.t.}  \nonumber\\
&=&\frac{g}{12\pi^2}\int_0^\infty \frac{k^3dk}{\exp[(\omega^*(k,T,\mu)-\mu)/T]\mp 1}\left.\frac{\partial \omega^*(k,T,\mu)}{\partial k}\right|_{T,\mu}+\mathrm{c.t.}\,\,. \label{fpid}
\end{eqnarray}
We note the presence of the term regarding $E_1$ in the resulting expression for the pressure, but not in that for the energy density.

The number density reads
\begin{eqnarray}
n=\frac{\langle N \rangle}{V}=-\frac{1}{V}\frac{\partial \ln Q_G}{\partial\alpha}=n_{\mathrm{id}} ,\label{fnb}
\end{eqnarray}
with
\begin{eqnarray}
n_{\mathrm{id}}=\frac{g}{2\pi^2}\int_0^\infty \frac{k^2dk}{\exp[(\omega^*(k,T,\mu)-\mu)/T]\mp 1}-\mathrm{c.t.} \,\, .
\end{eqnarray}
Again, the contribution from the chemical potential dependence of quasiparticle mass in the ideal gas term and that from $E_0$ term cancel out each other if $B$ satisfies
\begin{eqnarray}
\frac{\partial B}{\partial\mu}= -\frac{g}{2\pi^2}\int_0^\infty \frac{k^2dk}{\omega^*(k,T,\mu)}\frac{1}{\exp[(\omega^*(k,T,\mu)-\mu)/T]\mp 1}m\frac{\partial m}{\partial \mu} + \mathrm{c.t.}\,\,. \label{go3}
\end{eqnarray}

We note that the resultant expressions for the thermodynamic quantities, namely, Eq.(\ref{energydensity}), Eq.(\ref{pressure}) and Eq.(\ref{fnb}), are thermodynamically as well as statistically consistent.
The reasons are twofold.
Firstly, the expressions for energy and particle density are in accordance with the conventional definition regarding ensemble average\footnote{This can be seen by comparing the r.h.s. of Eq.(\ref{energydensity}) and Eq.(\ref{fnb}) against Eq.(\ref{ensembleavg}).}, while they can also been conveniently expressed in standard form as derivatives of the grand partition function, as emphasized by other authors~\cite{eos-quasiparticle-14,eos-quasiparticle-03}.
Moreover, from the viewpoint of statistical physics, those ensemble averages are meaningful, only when one can match those quantities, term by term, to the first law of thermodynamics~\cite{book-statistical-mechanics-pathria}.
In this context, thermodynamical consistency is guaranteed.
Subsquently, any other thermodynamical quantities can be then derived from a thermodynamic potential, which summarizes all the constitutive properties of a body that thermodynamics represents.
Now, it is not difficult to see that the second requirement is indeed achieved by evaluating the total derivative of $q=\ln Q_G$, to be specific, one can readily verify that
\begin{eqnarray}
dq=-\langle N \rangle d\alpha - \langle E \rangle d\beta - \beta p dV  .
\end{eqnarray}
By comparing the above expression with the first law of thermodynamics, namely,
\begin{eqnarray}
d\langle E\rangle=T dS - p dV + \mu d\langle N\rangle  .
\end{eqnarray}
it is inferred that
\begin{eqnarray}
&&\beta = \frac{1}{k_{B}T} ,  \nonumber \\
&&\alpha = -\frac{\mu}{k_{B}T} ,  \nonumber\\
&&q+\alpha N +\beta E = \frac{S}{k_{B}} .\label{eentropy}
\end{eqnarray}
Since the first law of thermodynamics holds, it is natural to expect that all thermodynamical quantities defined through the above procedure automatically satisfy any thermodynamical relations, such as:
\begin{eqnarray}
\epsilon \equiv \frac{E}{V} = T\left.\frac{\partial p}{\partial T}\right|_{V,\mu} - p +\mu n ,
\end{eqnarray}
which is frequently discussed in the literature.

As discussed above, for the case of finite density, $B$ has to satisfied both Eq.(\ref{go2}) and Eq.(\ref{go3}) simultaneously, which is not equivalent to Eq.(\ref{go0a}).
In fact, the symmetry of second derivatives for Eq.(\ref{go2}) and Eq.(\ref{go3}) implies the following integro-differential equation:
\begin{equation}
\begin{aligned}
\llangle m\frac{\partial m}{\partial T} \rrangle_- = \llangle m\frac{\partial m}{\partial \mu}\rrangle_+ , \label{gozero}
\end{aligned}
\end{equation}
where
\begin{equation}
\begin{aligned}
\llangle O \rrangle_- = \int_0^\infty k^2dk\left\{\frac{\exp[(\omega^*-\mu)/T]}{(\exp[(\omega^*-\mu)/T]\mp 1)^2 T}-\mathrm{c.t.}\right\}O(k) , \\
\llangle O \rrangle_+ = \int_0^\infty k^2dk\left\{\frac{\exp[(\omega^*-\mu)/T](\omega^*-\mu)}{(\exp[(\omega^*-\mu)/T]\mp 1)^2 T^2}+\mathrm{c.t.}\right\}O(k) . \label{gozero2}
\end{aligned}
\end{equation}
For the most general cases, the particle mass is a function of momentum, $m=m(k,T,\mu)$, and therefore $B$ is actually a functional of $m$ besides a function of $T$ and $\mu$, and derivatives in equations such as Eq.(\ref{go0a}) should be understood as functional derivatives.
In the present study, we assume for simplicity that for an anti-particle $\bar{X}$, $m_{\bar{X}}(k,T,-\mu)=m_X(k,T,\mu)\equiv m(k,T,\mu)$, and again, ``$\mathrm{c.t.}$" indicates the counter term due to the contributions from anti-particles, they are obtained from the foregoing term by substituting $\mu\rightarrow -\mu$ and $X\rightarrow \bar{X}$.
In what follows we will discuss two special solutions of Eq.(\ref{gozero}).

\subsection{III The momentum independent solution}

Let us first consider the case where the quasiparticle mass is only a function of temperature and chemical potential, $m=m(T,\mu)$.
Then both $m$ and its derivatives can be moved out of the integrals with respect to $k$, and therefore, Eq.(\ref{gozero}) gives
\begin{equation}
\begin{aligned}
\frac{\partial m}{\partial T}\llangle 1 \rrangle_- = \frac{\partial m}{\partial \mu}\llangle 1 \rrangle_+ \label{go9} ,
\end{aligned}
\end{equation}
or,
\begin{equation}
\begin{aligned}
\frac{\partial m}{\partial T}\frac{\partial}{\partial \mu}\left(\left.\frac{\partial p_{\mathrm{id}}}{\partial m}\right|_{T,\mu} \right)
=\frac{\partial m}{\partial \mu}\frac{\partial}{\partial T}\left(\left.\frac{\partial p_{\mathrm{id}}}{\partial m}\right|_{T,\mu}\right) . \label{go9b}
\end{aligned}
\end{equation}
when expressed in terms of $p_{\mathrm{id}}$ of Eq.(\ref{fpid}).
By summing both sides of the above equation to the Maxwell relation of the ideal gas,
\begin{equation}
\begin{aligned}
\frac{\partial}{\partial\mu}\left(\left.\frac{\partial p_{\mathrm{id}}}{\partial T}\right|_{m,\mu}\right)
=\frac{\partial}{\partial T}\left(\left.\frac{\partial p_{\mathrm{id}}}{\partial \mu}\right|_{m,T}\right) ,
\end{aligned}
\end{equation}
and taking into account Eq.(\ref{go2}) and (\ref{go3}), one recovers
\begin{equation}
\begin{aligned}
\frac{\partial s}{\partial \mu}=\frac{\partial^2 p}{\partial T\partial \mu} =\frac{\partial^2 p}{\partial \mu\partial T}=\frac{\partial n}{\partial T} .\label{eq7ref-eos-latt-11}
\end{aligned}
\end{equation}
This is a Maxwell relation, precisely Eq.(7) of Ref.\cite{eos-latt-11}, which was subsequently used to determine the flow equation for the running coupling constant.
Alternatively, from our viewpoint, Eq.(\ref{go9}) is a condition to determine the particle mass $m(T,\mu)$.
It is not difficult to see that Eq.(\ref{go9}) can be formally solved by using the method of characteristics.
As shown in Appendix, its solution consists of characteristic curves for given $m$ satisfying
\begin{eqnarray}
\frac{d \mu}{d T}=-\frac{\llangle 1 \rrangle_+}{\llangle 1 \rrangle_-} . \nonumber
\end{eqnarray}
One may make use of the lattice data at zero chemical potential as the boundary condition.
Then again, one may simply solve $m(T,\mu)$ by carrying out numerical integral from the $\mu=0$ boundary onto the $T-\mu$ plane where $\mu\ne 0$.

\subsection{IV A special momentum dependent solution}

In general, as the solution of Eq.(\ref{gozero}), the quasiparticle mass is a function of $k$, $T$ and $\mu$.
For this case, we only discuss a special solution which possesses a rather simple form.
It is obtained by assuming the integrands on the both sides are the same. In other words,
\begin{equation}
\begin{aligned}
\left\{\frac{\exp[(\omega^*-\mu)/T]T}{(\exp[(\omega^*-\mu)/T]\mp 1)^2}-\mathrm{c.t.}\right\}\frac{\partial m}{\partial T} =\left\{\frac{\exp[(\omega^*-\mu)/T](\omega^*-\mu)}{(\exp[(\omega^*-\mu)/T]\mp 1)^2}+\mathrm{c.t.}\right\}\frac{\partial m}{\partial \mu} . \label{godown}
\end{aligned}
\end{equation}
Since $\omega^*$ is involved in the above equation, the resultant particle mass is indeed a function of $k$.
Then again, the above equation can be solved by using the method of characteristics, and its solution consists of characteristic curves for given $\omega^*$.

In particular, if the contributions from anti-particles are insignificant, namely, $\mu \gg 1$, Eq.(\ref{godown}) can be further simplified to
\begin{eqnarray}
\frac{\partial m}{\partial \mu} =\frac{T}{(\omega^*(k,T,\mu)-\mu)}\frac{\partial m}{\partial T} ,  \label{goup}
\end{eqnarray}
which possesses the following analytic solution (see also the Appendix)
\begin{eqnarray}
m=f\left(\frac{T\omega^*}{\omega^*-\mu}\right) , \label{go7b}
\end{eqnarray}
where $f(T)\equiv m(T,\mu=0)$ is determined by the boundary condition.

\section{V Numerical Results}

\begin{figure}
\begin{tabular}{cc}
\begin{minipage}{250pt}
\centerline{\includegraphics[width=250pt]{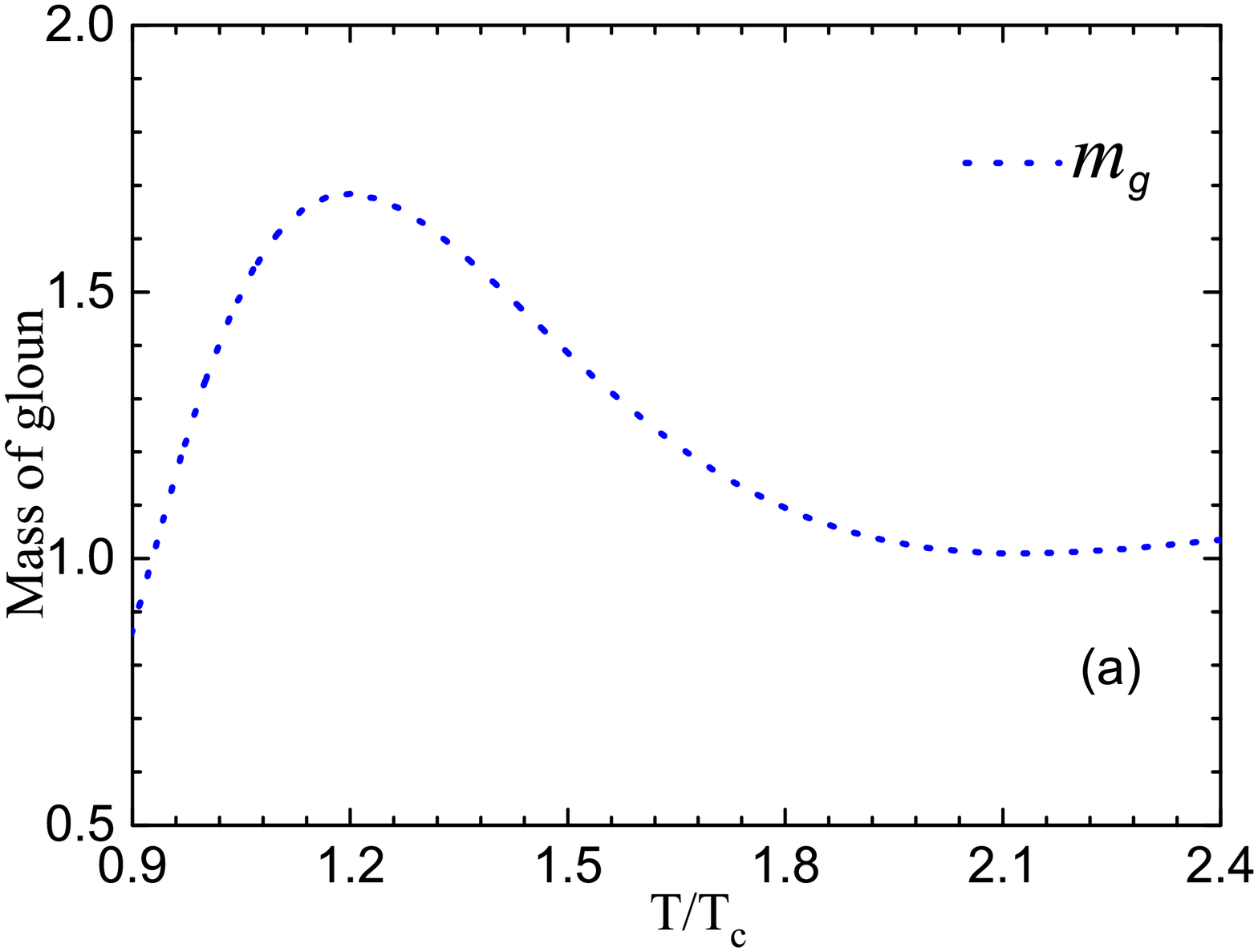}}
\end{minipage}
&
\begin{minipage}{250pt}
\centerline{\includegraphics[width=250pt]{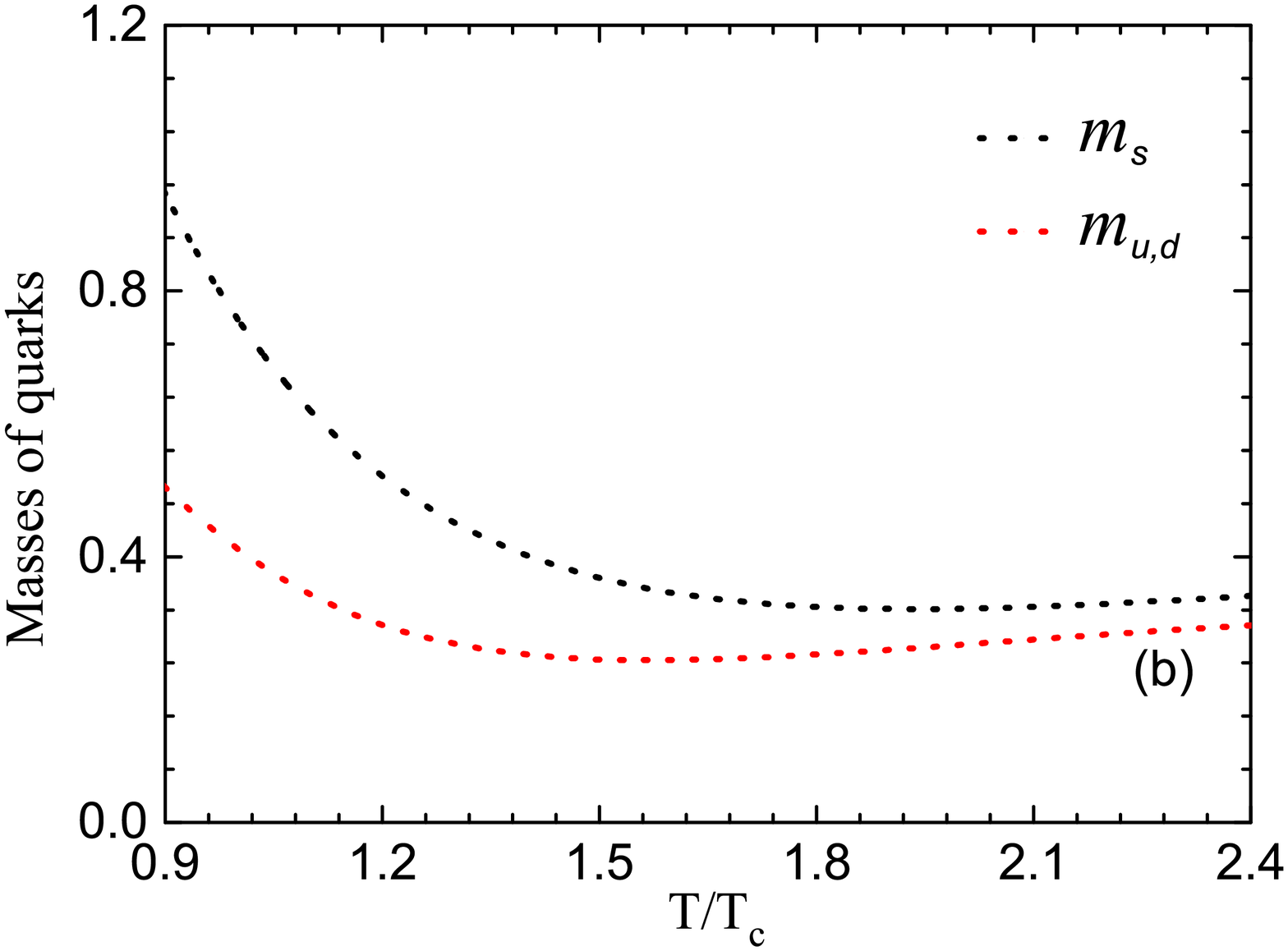}}
\end{minipage}
\\
\end{tabular}
\caption{(Color online) The resultant temperature dependent quasiparticle mass for glouns, light and strange quarks at zero chemical potential.}
 \label{fitmt}
\end{figure}

Now, we are in a position to present the numerical results and to compare them to the recent lattice data for $N_f=2+1$ favor QCD system~\cite{lattice-12,lattice-14,lattice-18,lattice-15,lattice-19}.
We first show the calculated thermodynamical quantities for the case of momentum dependent quasiparticle mass.
Here, the free parameters are the effective masses of gluons, of light as well as strange quarks as functions of temperature at zero chemical potential, and a constant related to $E_1$.
Once they are determined, one may evaluate all thermodynamical quantities such as energy density, pressure, and entropy density at zero as well as finite baryon density.
In addition, we also calculate the trace anomaly, sound velocity, and the particle number susceptibility defined as
\begin{eqnarray}
\chi^{ab}_2=\frac{T}{V}\frac{1}{T^2}\left.\frac{\partial^2 \ln Q_G(T,\mu_u,\mu_d,\mu_s)}{\partial\mu_a\partial\mu_b}\right|_{\mu_a=\mu_b=0} .
\end{eqnarray}

\begin{figure}
\begin{tabular}{cc}
\begin{minipage}{250pt}
\centerline{\includegraphics[width=250pt]{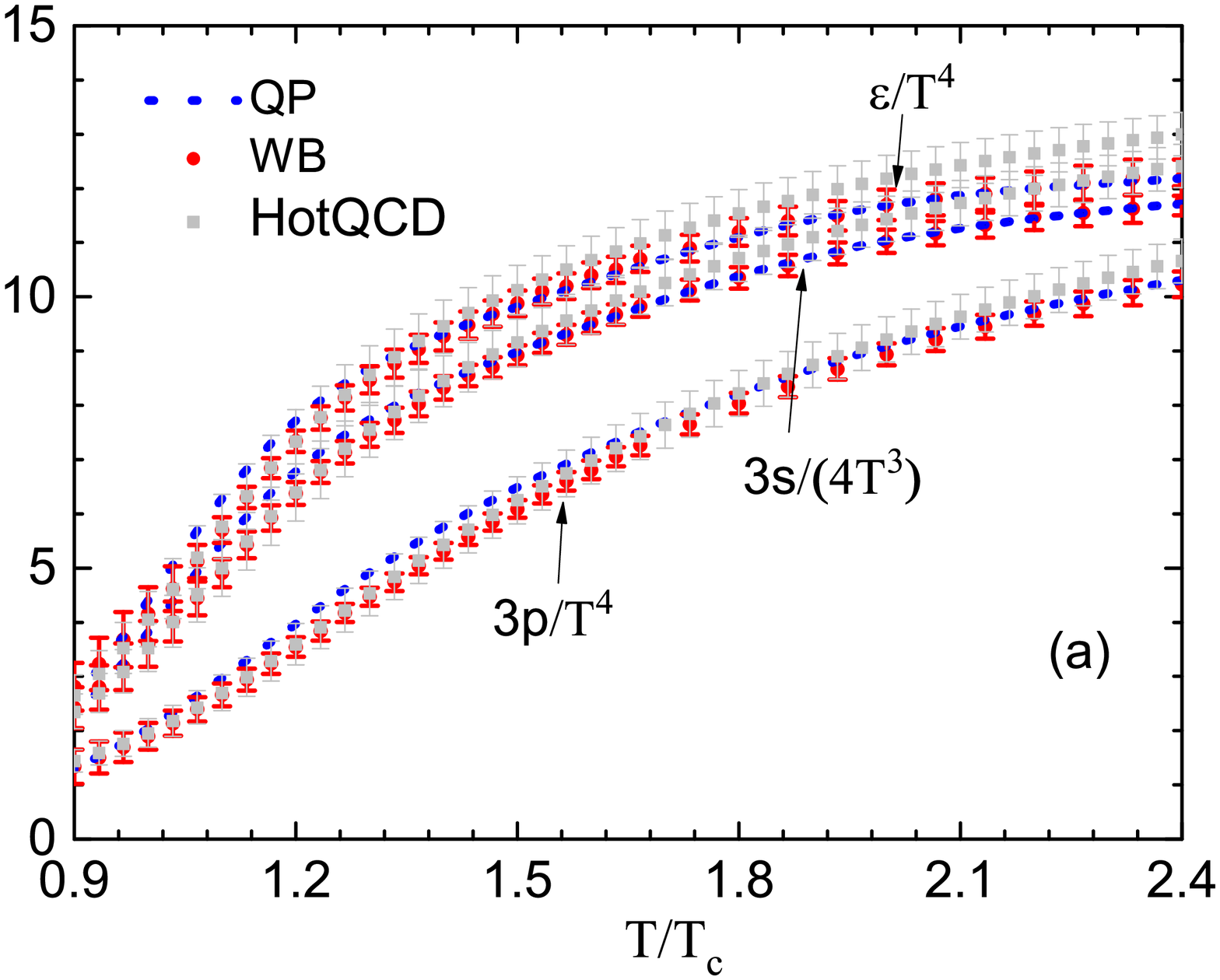}}
\end{minipage}
&
\begin{minipage}{250pt}
\centerline{\includegraphics[width=250pt]{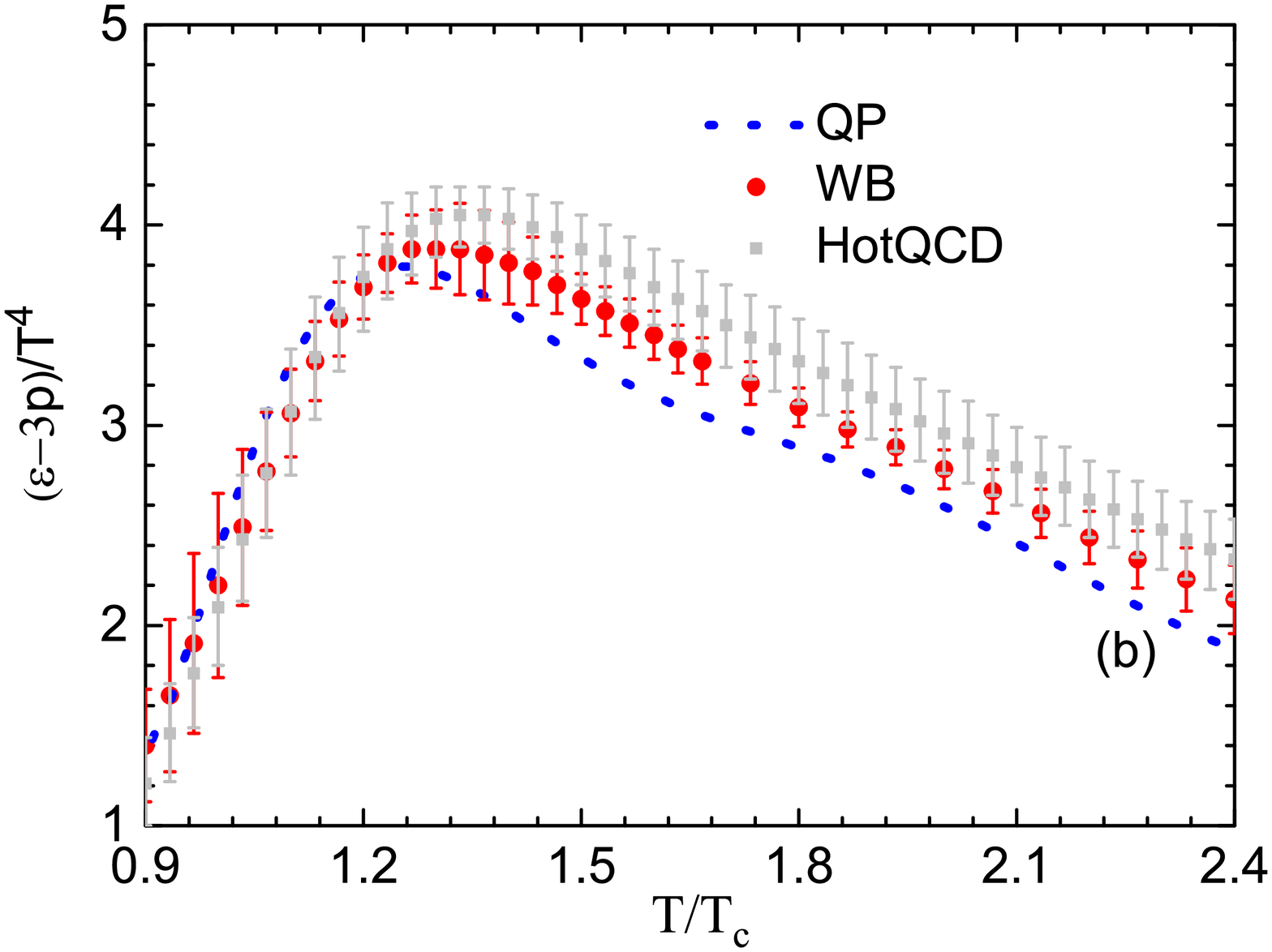}}
\end{minipage}
\\
\begin{minipage}{250pt}
\centerline{\includegraphics[width=250pt]{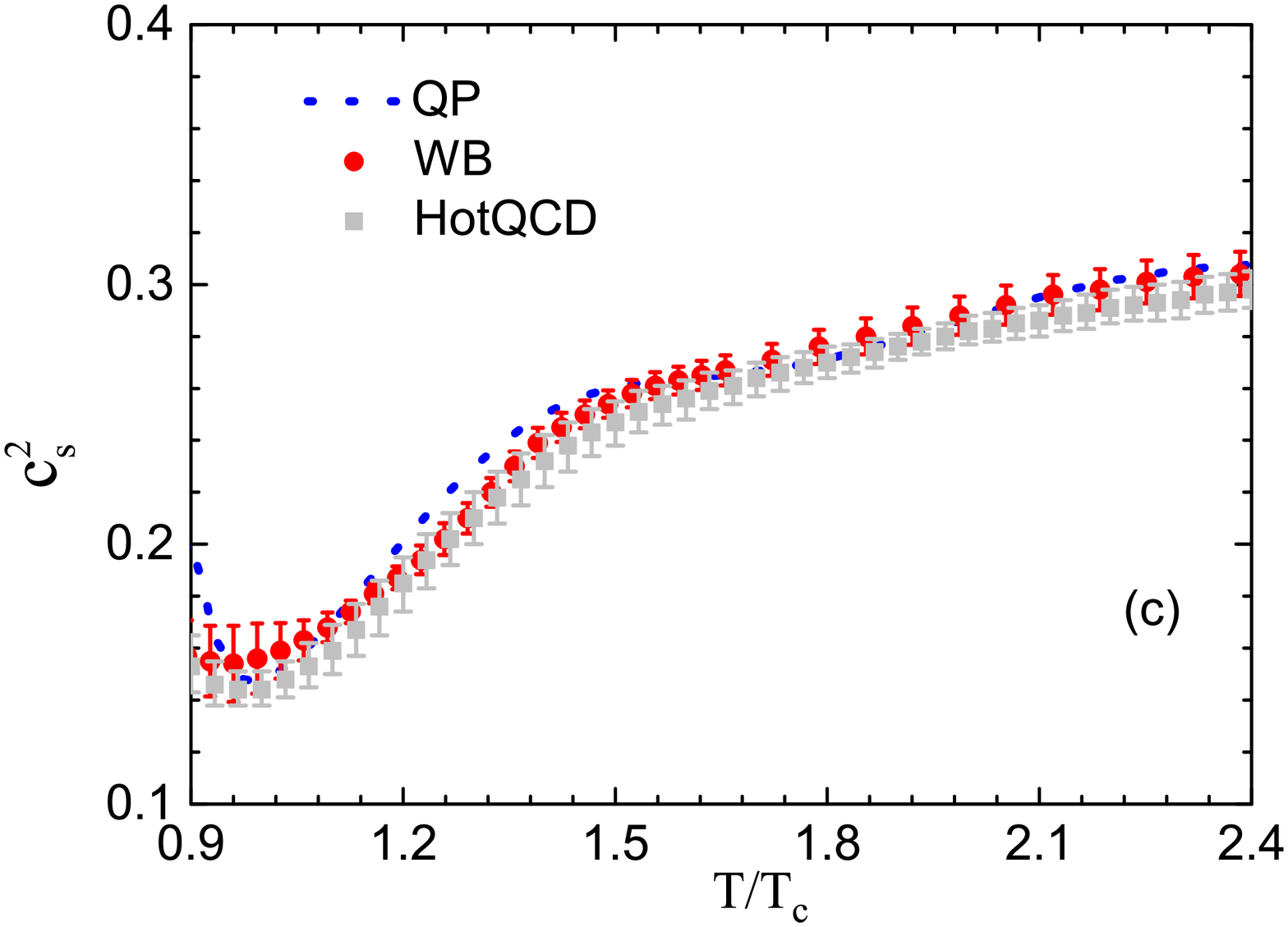}}
\end{minipage}
&
\begin{minipage}{250pt}
\centerline{\includegraphics[width=250pt]{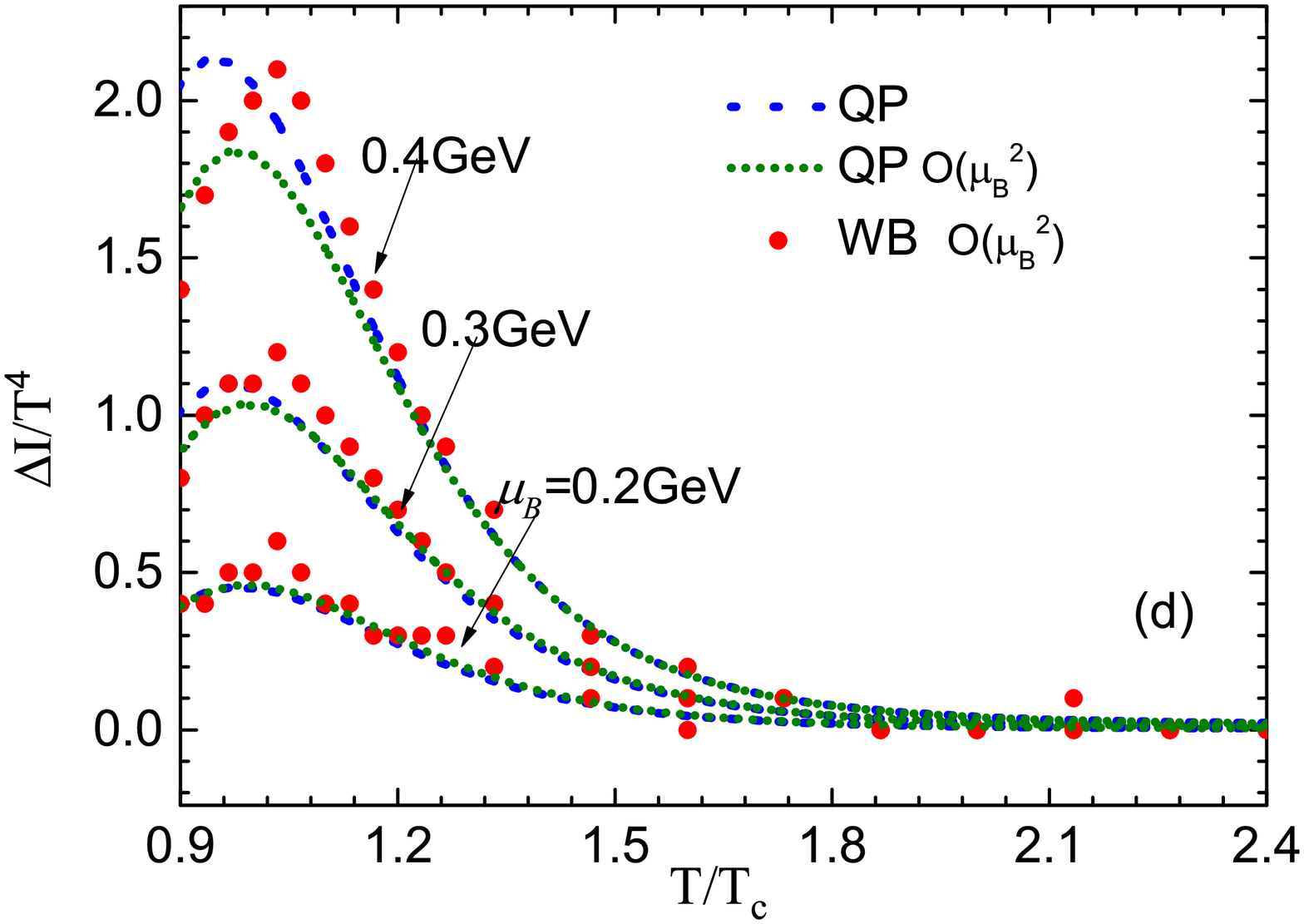}}
\end{minipage}
\end{tabular}
\caption{(Color online)
The calculated thermodynamical quantities for both vanishing and finite baryon chemical potential.
The thermodynamical quantities obtained by the present model is shown in dotted blue curves.
The calculated results truncated in terms of $\frac{\mu}{T}$ up to second are shown in dotted green curves.
They are compared with those of lattice QCD calculations the Wuppertal-Budapest~\cite{lattice-12,lattice-14} and HotQCD~\cite{lattice-18,lattice-15,lattice-19} Collaborations, indicated by filled red circles and grey squares (with error bars when it applies) respectively.
(a) and (b): the results of entropy density, energy density, pressure and trace anomaly at zero baryon chemical potential.
(c): calculated speed of sound.
(d): trace anomaly for different values of chemical potential.
}
\label{fit2latticeA}
\end{figure}

In particular, the relavant quantities $\chi^{B}_2$ and $\chi^{L}_2$ in the present model~\cite{lattice-12} read:
\begin{eqnarray}
\chi^{B}_2=\frac{1}{9}\left[\chi^u_2+\chi^d_2+\chi^s_2+2\chi^{us}_{11}+2\chi^{ds}_{11}+2\chi^{ud}_{11}\right]=\frac{1}{9}\left[2\chi^u_2+\chi^s_2\right] ,
\end{eqnarray}
and
\begin{eqnarray}
\chi^{L}_2=\frac{1}{9}\left[\chi^u_2+\chi^d_2+2\chi^{ud}_{11}\right]=\frac{2}{9}\chi^u_2 . \label{eq33}
\end{eqnarray}
The x-axis of the plots are chosen to be $T/T_c$, where the value for the transition temperature $T_c=0.15 $GeV is taken~\cite{lattice-14,lattice-15,lattice-09,lattice-08}.
Then all these results are compared to those obtained by the lattice QCD calculations by Wuppertal-Budapest~\cite{lattice-12,lattice-14} as well as HotQCD collaborations~\cite{lattice-18,lattice-15,lattice-19}.

The parameters of the present approach are determined as follows.
Firstly, the lattice data~\cite{lattice-14} on particle susceptibility of light $\chi_2^L$ quarks is used to determine the quasiparticle mass of light quarks at vanishing chemical potential.
Subsequently, the quasiparticle mass of strange quark as a function of temperature is determined by the particle susceptibility regarding baryon chemical potential $\chi_2^B$.
Then the gluon mass is used to fit the energy density for $n_B=0$.
Also, to compare the results between Eq.(\ref{go9}) and Eq.(\ref{godown}), we assume that the quasiparticle mass is momentum independent at $\mu=0$\footnote{This is a simplifying assumption, and it may be not valid in general. A more realistic approach is to accommodate the existing results regarding the momentum dependence of parton mass.}, so that both equations are solved by using the same boundary condition.
Finally, $E_1$ is tuned to further improve the pressure as a function of temperature at zero baryon density. 
The resultant particle masses at $\mu_q=0$ are show in Fig.\ref{fitmt}, the constant of integration for the bag constant is taken to be $B(T_c, \mu=0) =0.12 \times T_c^{4}$, and the value of $E_1$ is found to be $\beta E_1/V=2.305 \times 10^{-4} $GeV$^3 $.
The particle mass at finite chemical potential is subsequently evaluated according to Eq.(\ref{godown}).

We note that, in principle, it seems to be more reasonable to adjust the model parameters to the lattice data of $\chi_2^u$ and $\chi_2^s$, instead of $\chi_2^L$ and $\chi_2^B$.
This is because the lattice results show that $\chi_2^B$ contains flavor correlations.
However, since our quasiparticle model does not take into account the contributions from mixed cumulant terms, such as $\chi_{11}^{ud}$ in Eq.(\ref{eq33}), it is found that in practice the proposed model calibration leads to a better fit to the existing lattice data.

\begin{figure}
\begin{tabular}{cc}
\begin{minipage}{250pt}
\centerline{\includegraphics[width=250pt]{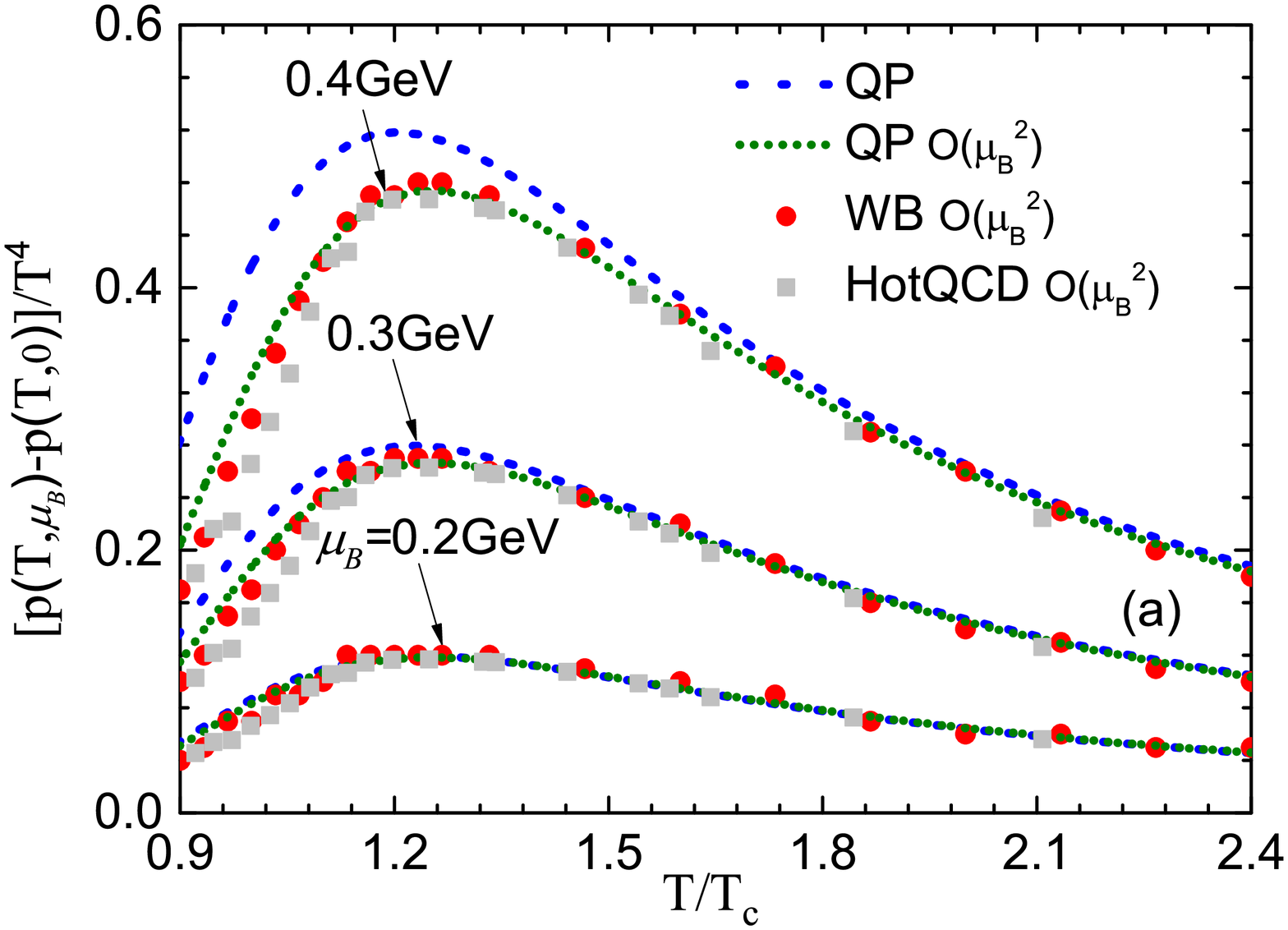}}
\end{minipage}
&
\begin{minipage}{250pt}
\centerline{\includegraphics[width=250pt]{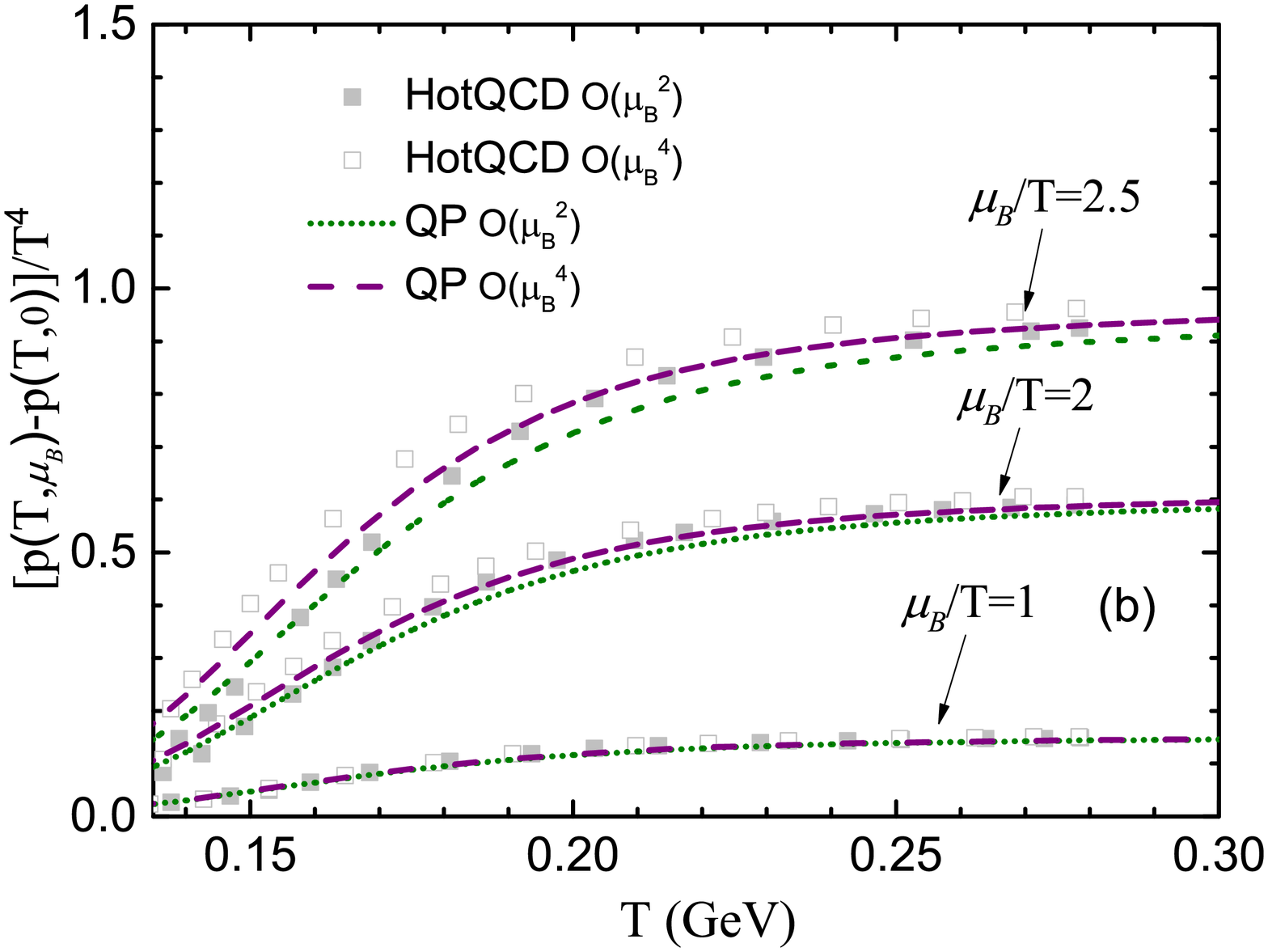}}
\end{minipage}
\\
\begin{minipage}{250pt}
\centerline{\includegraphics[width=250pt]{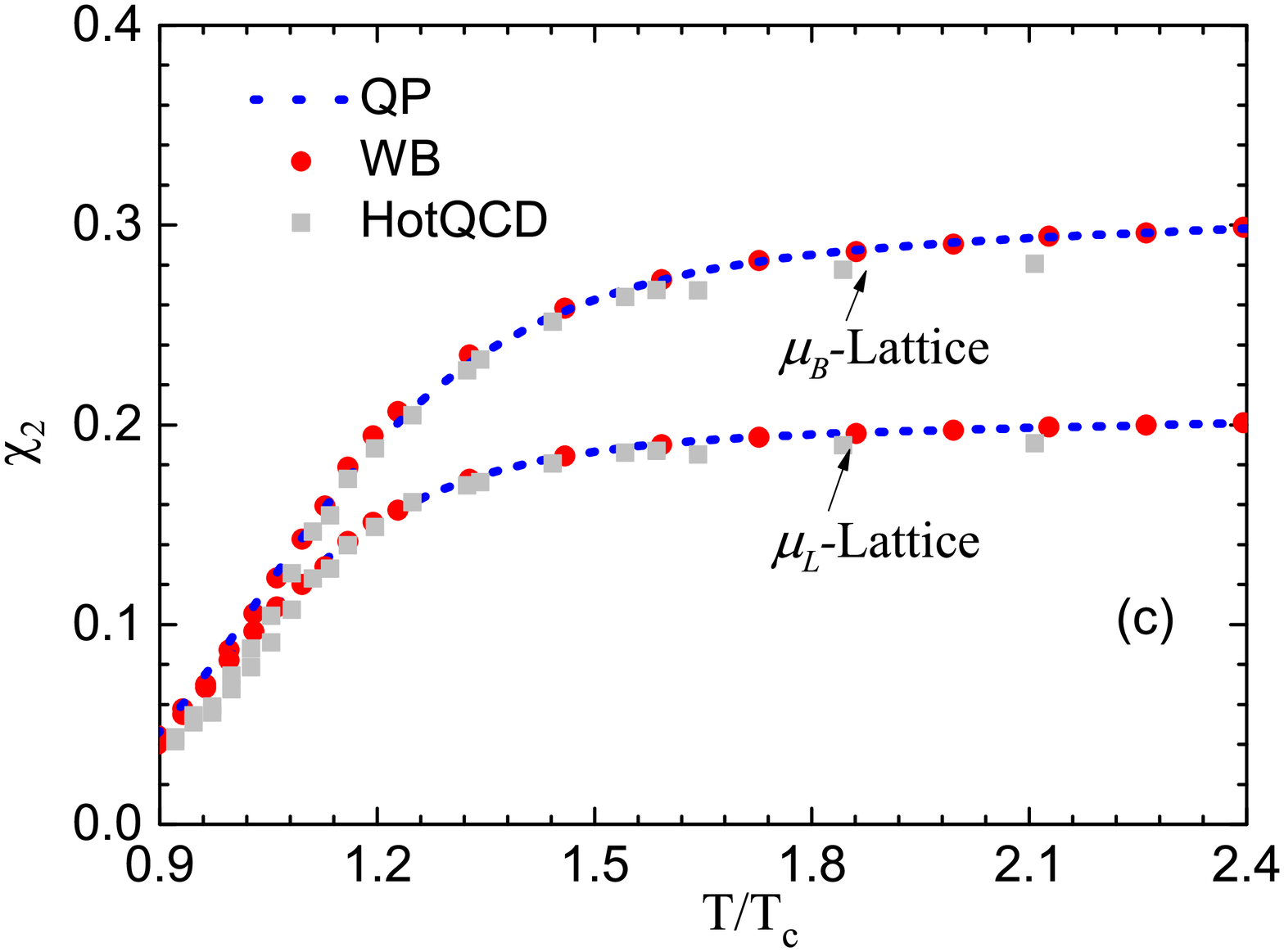}}
\end{minipage}
&
\begin{minipage}{250pt}
\centerline{\includegraphics[width=250pt]{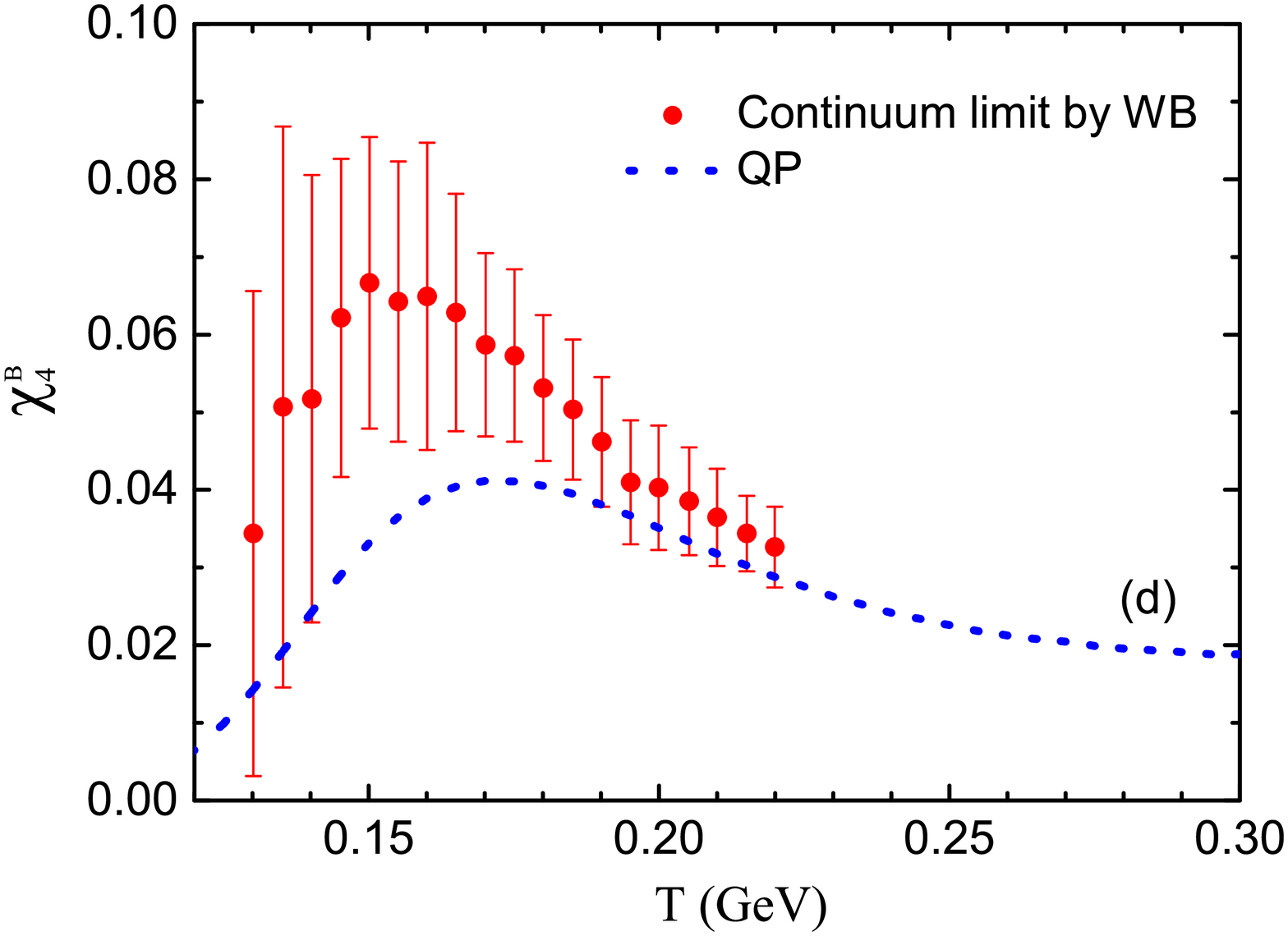}}
\end{minipage}
\end{tabular}
\caption{(Color online)
The calculated thermodynamical quantities for both vanishing and finite baryon chemical potential.
The thermodynamical quantities obtained by the present model is shown in dotted blue curves.
The calculated results truncated in terms of $\frac{\mu}{T}$ up to second and fourth order are shown in dotted green and dashed purple curves respectively.
They are compared with those of lattice QCD calculations the Wuppertal-Budapest~\cite{lattice-12,lattice-14} and HotQCD~\cite{lattice-18,lattice-15,lattice-19} Collaborations, indicated by filled red circles and grey squares (with error bars when it applies) respectively.
(a) and (b): the difference of pressure for given $\mu_B$ or $\mu_B/T$ as a function of temperature.
The calculations have been carried out by using different truncations and the results are compared against corresponding lattice data.
(c) and (d): the second and fourth order cumulants of particle number fluctuations, $\chi_2$ and $\chi_4$.
}
\label{fit2latticeB}
\end{figure}

The resultant thermodynamic quantities are presented in Fig.\ref{fit2latticeA} and Fig.\ref{fit2latticeB}.
One observes that, overall, a reasonably good agreement is achieved, especially for quark number susceptibility, besides the energy density, entropy density, and pressure.
It is also worth pointing out that in our present approach, we did not introduce any renormalization for the degeneracy factor, which is adopted as an additional free parameter by some of the quasiparticle approaches.
The only discrepancies are observed for the quantities associated with the first and second derivative of the grand partition function for the region where $T<T_c$ .
For instance, the pressure difference is related to the expansion in terms of $\mu/T$.
Therefore the deviation becomes larger for smaller temperature.
It is probably related to the peak of $\chi_4$ at $T_c$~\cite{lattice-16} which has not been appropriately considered in the present study.
As explained above, the fit was only carried out regarding the $\chi_2$ lattice data.
Since the lattice QCD results were obtained by a Taylor expansion in terms of $\frac{\mu}{T}$, it is thus meaningful to show our results also truncated to the corresponding order when comparing to them.
This is shown in Fig.\ref{fit2latticeB} (c) and (d).
It is noted when we evaluate the pressure difference expanded up to the order of $\left(\frac{\mu}{T}\right)^2$, the calculated curve stays closer to the lattice results, as expected.
The is shown by the dotted green curves in Fig.\ref{fit2latticeB} (c).
But since the present quasiparticle model does not consider any contribution from the mixed second order derivative such as $\chi^{ud}_{11}$, it is merely understood as a result of appropriate parameterization.
For the same reason, the results on fourth-order cumulant $\chi_4^B$ presents more substantial discrepancies.
Probably due to a similar reason, some small deviation is also found for the calculated sound velocity as a function of temperature.
However, by adjusting the gluon mass in the region of temperature $T\sim T_c$, we were able to reproduce the behavior of the sound speed which increases again as the temperature reaches the sector associated with the hadronic resonance gas.
From a practical viewpoint, these difference can also be amended by manually connecting the quasiparticle EoS to that of the hadronic resonance gas model.

\begin{figure}
\begin{tabular}{cc}
\begin{minipage}{250pt}
\centerline{\includegraphics[width=250pt]{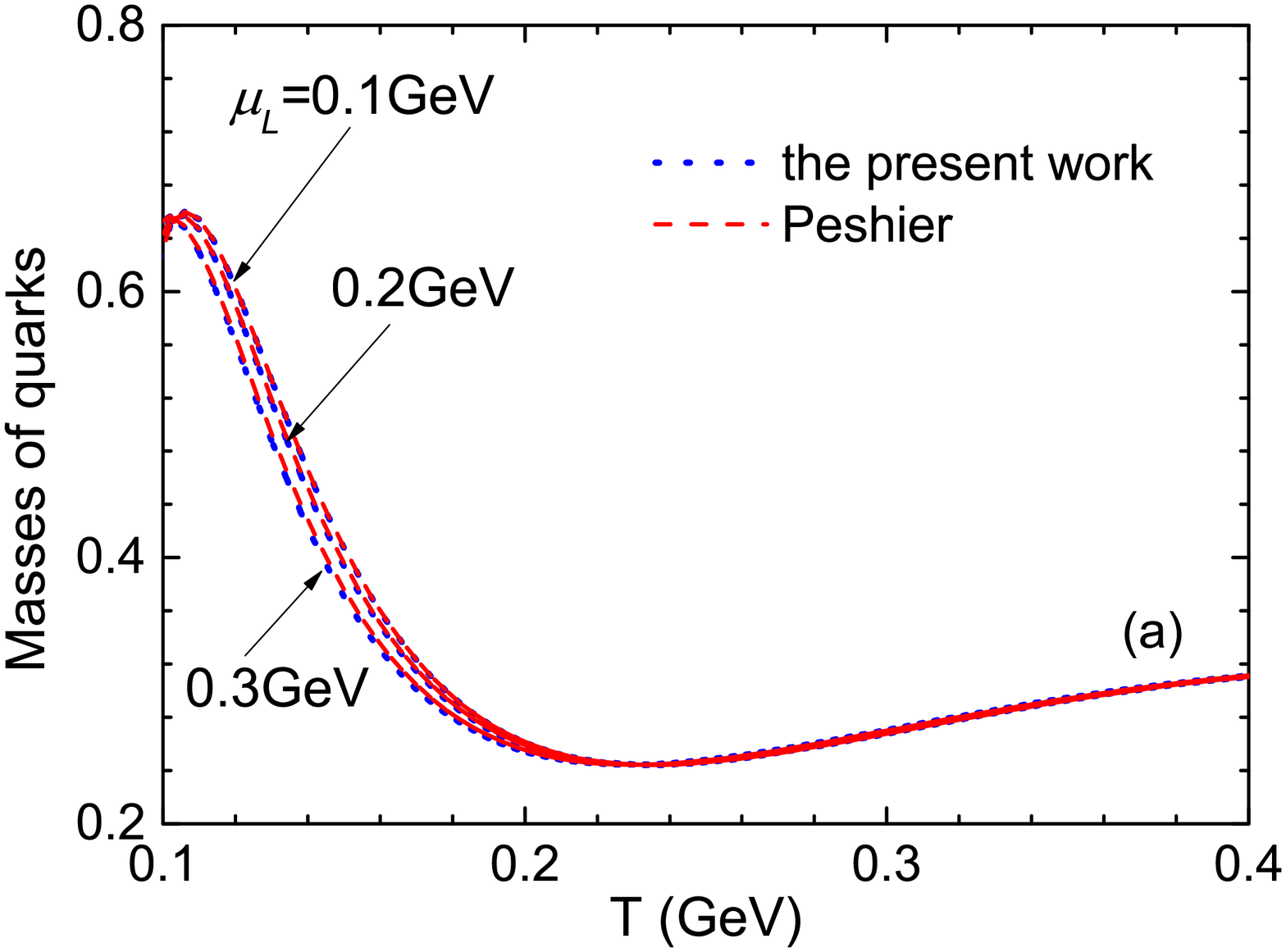}}
\end{minipage}
&
\begin{minipage}{250pt}
\centerline{\includegraphics[width=250pt]{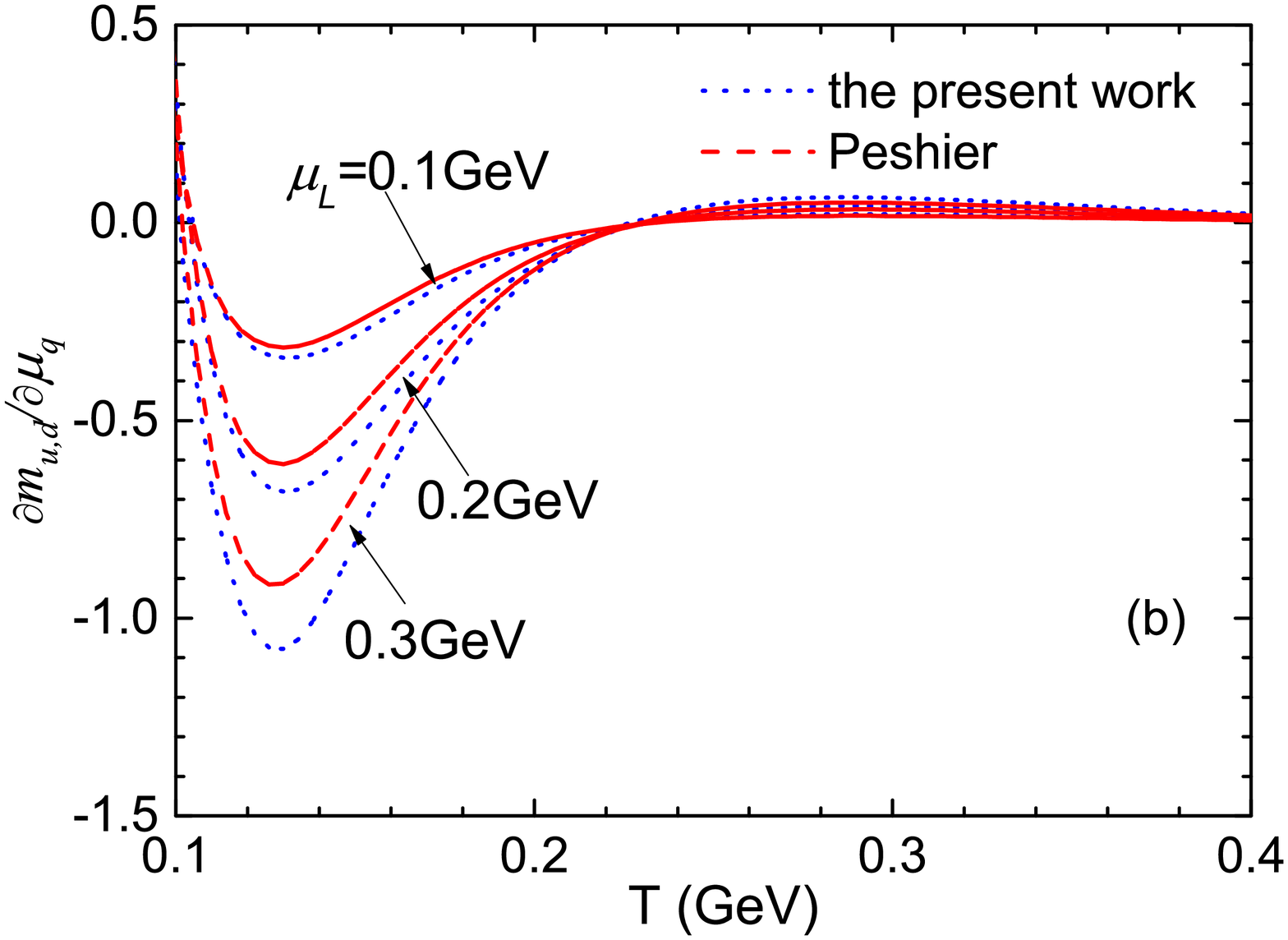}}
\end{minipage}
\\
\begin{minipage}{250pt}
\centerline{\includegraphics[width=250pt]{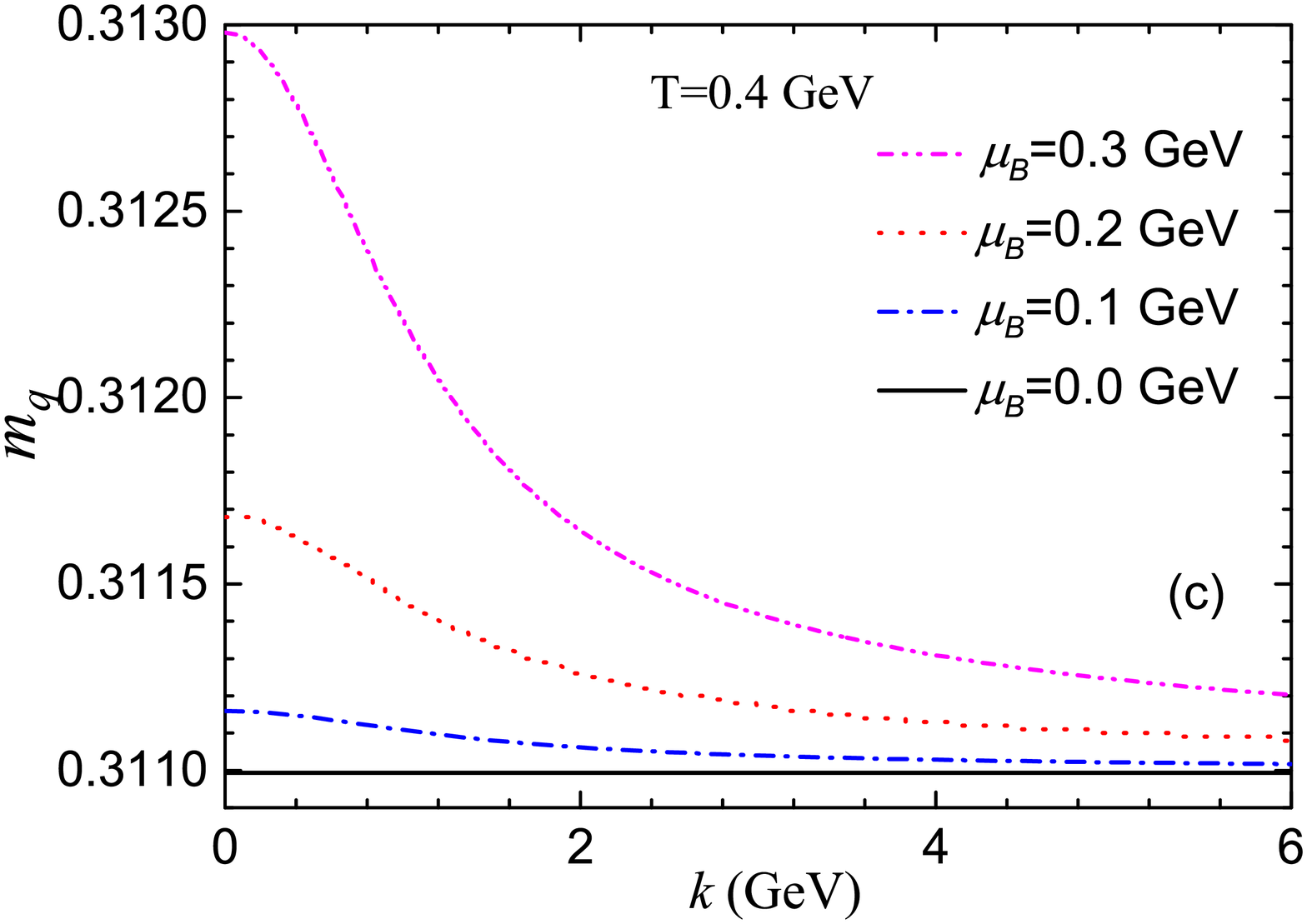}}
\end{minipage}
&
\begin{minipage}{250pt}
\centerline{\includegraphics[width=250pt]{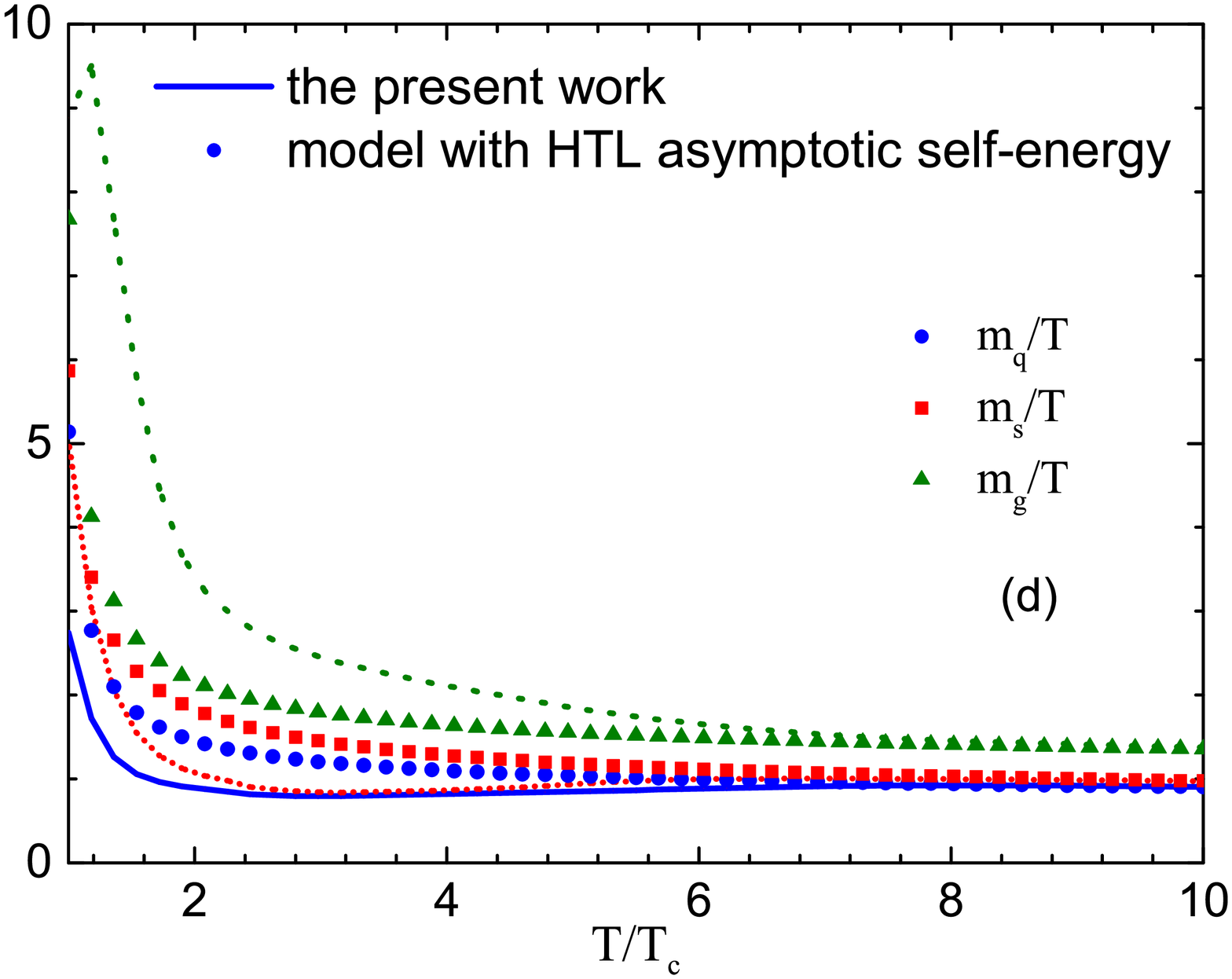}}
\end{minipage}
\end{tabular}
\caption{(Color online) 
(a) and (b): the calculated quasiparticle mass of light quarks and its derivative as functions of temperature for different baryon chemical potentials, obtained by solving Eq.(\ref{godown}), in comparison with those by solving Eq.(\ref{go9}), the latter is equivalent to the approach by Peshier {\it et al}.~\cite{eos-quasiparticle-13}.
(c): the quasiparticle mass of light quarks as a function of momentum for the solution discussed in this work.
(d): the calculated asymptotic behavior of quasiparticle masses, in comparison with a model~\cite{eos-latt-16} inspired by the gauge-independent hard thermal/dense loop (HTL) calculations.}
\label{massfunction}
\end{figure}

To order to compare two different solutions discussed in the previous section, we solve Eq.(\ref{go9}) and Eq.(\ref{godown}) respectively but fitting to the given boundary condition at $n_B=0$ defined by the lattice data.
The corresponding results are shown in Fig.\ref{massfunction}, where the obtained particle masses are presented as a function of temperature.
In the first case, since the mass is also a function of momentum $k$, the presented results are average values evaluated by using the same weight on the r.h.s. of Eq.(\ref{go2}) or (\ref{go3}).
Numerically, one finds that the particles masses from two different schemes are quite close to each other.
Though it seems to be a somewhat a surprising result, we understand that it could be merely owing to that both approaches are tuned to reproduce the lattice data and the fact the numerically obtained momentum dependence of quasiparticle mass is not strong at all.
The latter is observed in Fig.\ref{massfunction} (c), which presents the obtained momentum dependence of quark masses for a given temperature but with different values of chemical potential.
It is observed that the quasiparticle mass decreases slightly but monotonically and converges to a given value as the momentum increases.
As the chemical potential increases, the dependence becomes stronger, though the overall dependence is not significant.

Last but not least, we show that the results obtained in the present approach are consistent with the established perturbative limit.
This is achieved by carrying out calculations by using the quasiparticle model proposed in Ref.~\cite{eos-latt-16} with the following forms for the quasiparticle masses
\begin{eqnarray}
m_{a}^{2}=m_{a0}^{2}+ \Pi_{a} ,
\end{eqnarray}
where $a = g, q, s$ and the quasiparticle self-energies adopt the asymptotic forms of the gauge-independent hard thermal/dense loop (HTL) calculations~\cite{qcd-HTL-01,book-thermo-field-theory-Bellac}:
\begin{eqnarray}
\Pi_{g}&&= \left(  \left[ 3+ \frac{N_f}{2} \right] T^2 + \frac{3}{2 \pi^{2}} \sum_f \mu_{f}^{2}  \right) \frac{G^2}{6}, \\
\Pi_{q}&&=2 m_{q0} \sqrt{\frac{G^2}{6} \left( T^2 + \frac{\mu_{q}^{2}}{\pi^{2}} \right)} + \frac{G^2}{3} \left( T^2 + \frac{\mu_{q}^{2}}{\pi^{2}} \right),\\
\Pi_{s}&&=2 m_{s0} \sqrt{\frac{G^2}{6} T^2} + \frac{G^2}{3} T^2 . \label{mHTL}
\end{eqnarray}
For the high temperature region, the coupling is taken to have the form of the perturbative running coupling at two-loop order
\begin{eqnarray}
G^2(T,\mu_q=0)=\frac{16 \pi^{2}}{\beta_0 \log \xi^2} \left[ 1- \frac{2 \beta_1}{\beta_0} \frac{\log \left(\log \xi^2 \right)}{\log \xi^2}\right] ,
\end{eqnarray}
with
\begin{eqnarray}
\beta_0&&=\frac{11 N_c - 2 N_f}{3},\\
\beta_1&&=\frac{34 N_{c}^{2} - 13 N_c N_f - 3 N_f /N_c }{6} ,
\end{eqnarray}
and
\begin{eqnarray}
\xi=\lambda \frac{T-T_s}{T_c}.
\end{eqnarray}
which regulates the infrared divergence of the running coupling.
For the parameters, the scale parameter and the temperature shift are chosen to be $\lambda=1.5$ and $T_s=0.15 T_c$.
This is done so that the model may adequately reproduce the recent lattice data~\cite{lattice-12,lattice-14} in the intermediate temperature region, while the remaining parameters are taken to be the same as used in literature~\cite{eos-quasiparticle-feg-1}.
The calculated asymptotic behavior of the quasiparticle mass is shown in Fig.\ref{massfunction} (d).
As described above, the quasiparticle masses at vanishing chemical potential are adjusted to reproduce the lattice data at the intermediate temperature.
We first interpolate the lattice data, and then make use of the obtained expression to evaluate the particle masses for the whole temperature range.
The interpolation is carried out by specifically requiring the asymptotic behavior in Eq.(\ref{mHTL}) is attained at the limit $T\rightarrow \infty$.
It is shown that our present approach is indeed consistent with the established perturbative limit.
Owing to Eq.(\ref{mHTL}), at very high temperature but physically relevant finite chemical potential, the limit established above does not change at all, which is also confirmed by the numerical calculations.

\section{VI Concluding remarks}

To summarize, in this work we study the thermodynamic consistency of the quasiparticle model and its implications on quasiparticle mass.
We have found new possible solutions that have not be explored before, and an essential characteristic of these solutions is that the quasiparticle mass is also a function of the momentum.
Consequently, thermodynamical quantities are actually {\it functionals} of particle mass, and in this case, the formulation concerning the derivatives with respect to $m$, such as $dB/dm$ on the l.h.s. of Eq.(\ref{go0a}), cease to be well defined.
As discussed in the previous sections, such momentum dependence of quasiparticle mass is not a free parameterization but is derived from the requirement of thermodynamical consistency.
In particular, we investigated one special solution, and find that it is consistent with the most recent lattice data.
In fact, the momentum dependent effective mass is a meaningful concept.
For instance, results on the gluon~\cite{qcd-RGZ-01,qcd-RGZ-02,qcd-RGZ-04,qcd-RGZ-05} and quark propagator~\cite{qcd-GZ-02} in terms of the Gribov-Zwanziger framework show that the resultant pole masses indeed are functions of momentum.
Also, other non-perturbative approaches such as the Schwinger-Dyson equation indicate that both gluon~\cite{qcd-DSE-02} and quark~\cite{qcd-DSE-03,qcd-DSE-04} dynamic masses are momentum dependent.
In particular, the concept of momentum dependent self-energy has been investigated by many authors in the context of quasiparticle model~\cite{eos-quasiparticle-17,eos-quasiparticle-18,eos-quasiparticle-19,eos-quasiparticle-20}.
Besides, we show that the scenario discussed previously by other authors~\cite{eos-latt-11,eos-quasiparticle-16,eos-latt-16,eos-latt-12} can be readily restored if one enforces that quasiparticle mass is only a function of temperature and chemical potential.
From our viewpoint, however, the derived ``flow equation" for the running coupling~\cite{eos-latt-11} can alternatively be written down as an equation in terms of the quasiparticle mass.
We also investigated a special solution where quasiparticle mass is a function of the momentum, by simply matching the integrants of the integro-differential equation.
By numerical calculations, we show that the difference between these different schemes are not very significant, once the lattice data at zero chemical potential is used as a constraint.

Partly inherited from most quasiparticle approaches, the present model does not naturally address the flavor off-diagonal correlations.
The latter subsequently leads to deviation from the lattice data in the transition region at fourth order and beyond.
Also, as the present model still show some discrepancy from the lattice data for the region $T<T_c$, it seems natural to smoothly connect the EoS in this region to that of hadronic resonance gas model.
In Ref.~\cite{sph-eos-1}, a critical point is implemented phenomenologically at finite baryon chemical potential.
Since the EoS plays an essential role in the hydrodynamic description of relativistic heavy-ion collisions~\cite{sph-review-1,sph-eos-2,sph-eos-3}, one can employ this scheme to study the properties of the system regarding the existence of the critical point, especially their particular consequences owing to the hydrodynamic evolution of the system.
Hopefully, some observables can be compared to the ongoing RHIC beam energy scan program~\cite{RHIC-star-bes-01,RHIC-star-bes-02,RHIC-star-bes-03,RHIC-star-bes-04}.
We plan to carry out a hydrodynamic study of the relevant quantities using the proposed EoS.

\section*{Acknowledgments}
We are thankful for valuable discussions with Bruno W. Mintz, Arlene C. Aguilar, Tereza Mendes, and Aritra Bandyopadhyay.
We gratefully acknowledge the financial support from
Funda\c{c}\~ao de Amparo \`a Pesquisa do Estado de S\~ao Paulo (FAPESP),
Funda\c{c}\~ao de Amparo \`a Pesquisa do Estado do Rio de Janeiro (FAPERJ),
Conselho Nacional de Desenvolvimento Cient\'{\i}fico e Tecnol\'ogico (CNPq),
and Coordena\c{c}\~ao de Aperfei\c{c}oamento de Pessoal de N\'ivel Superior (CAPES).
A part of the work was developed under the project INCTFNA Proc. No. 464898/2014-5.
This research is also supported by the Center for Scientific Computing (NCC/GridUNESP) of the S\~ao Paulo State University (UNESP).

\section*{Appendix}

In this section, we show how the solutions of Eq.(\ref{go9}) and Eq.(\ref{goup}) are obtained.
As a matter of fact, the procedure to solve the above equations is very similar, while the latter is slightly more complicated.
Therefore, in what follows, we explicitly derive the solution of Eq.(\ref{goup}) and briefly discuss how that of Eq.(\ref{go9}) is obtained.
One first rewrites Eq.(\ref{goup}) by defining
\begin{eqnarray}
w = \omega^*-\mu
\end{eqnarray}
Since $m=\sqrt{(w+\mu)^2-k^2}$, considering $k$ merely as a parameter in $m=m(k,T,\mu)$, and $(w+\mu)$ as an intermediate variable, one has
\begin{eqnarray}
\frac{\partial m}{\partial \mu}&=&\frac{\partial m}{\partial (w+\mu)}\frac{\partial{(w+\mu)}}{\partial \mu} , \nonumber \\
\frac{\partial m}{\partial T}&=&\frac{\partial m}{\partial (w+\mu)}\frac{\partial{(w+\mu)}}{\partial T}=\frac{\partial m}{\partial (w+\mu)}\frac{\partial{w}}{\partial T} . \nonumber
\end{eqnarray}
Thus Eq.(\ref{goup}) implies
\begin{eqnarray}
\frac{\partial{(w+\mu)}}{\partial \mu}  = \frac{T}{w}\frac{\partial{w}}{\partial T} ,
\end{eqnarray}
or equivalently,
\begin{eqnarray}
w\frac{\partial{w}}{\partial \mu}  - T\frac{\partial{w}}{\partial T}+w=0 ,
\end{eqnarray}
whose solution can be obtained by using the method of characteristics~\cite{book-methods-mathematical-physics-01}.
To be specific, the above partial different equation can be fit into the formal form
\begin{eqnarray}
a(\mu,T,w)\frac{\partial w}{\partial \mu}+b(\mu,T,w)\frac{\partial w}{\partial T}=c(\mu,T,w) ,\label{ffchar}
\end{eqnarray}
with
\begin{eqnarray}
a(\mu,T,w)&=&w ,\nonumber \\
b(\mu,T,w)&=&-T ,\nonumber \\
c(\mu,T,w)&=&-w .
\end{eqnarray}
whose formal solution is the surface, defined by $f(\mu,T,w)=w-w(\mu,T)=0$, tangent to the vector field $(a(\mu,T,w),b(\mu,T,w),c(\mu,T,w))$, namely,
\begin{eqnarray}
\frac{d\mu}{w}=\frac{dT}{-T}=\frac{dw}{-w} .
\end{eqnarray}
As it contains two independent equtions, one may conveniently select
\begin{eqnarray}
d(\mu+w)&=&0 ,\nonumber
\end{eqnarray}
and
\begin{eqnarray}
d\left[\ln\left(\frac{w}{T}\right)\right]&=&d\left(\frac{w}{T}\right)=0 .\nonumber
\end{eqnarray}
This indicates that, for any function $F(u,v)$, the desired solution $w$ satisfies
\begin{eqnarray}
F\left(\frac{w}{T},(w+\mu)\right)=0 .
\end{eqnarray}
Now, as disscuss in the above text, the solution of the equation is determined by the boundary condition at $\mu=0$, where $m(k,T,\mu=0)\equiv f(T)$.
In other words, the form of $F$ shall be determined by the boundary condition.
If one defines $F_0(u,v)\equiv F(\mu=0)$, it is readily to verify that\footnote{It is in fact one of many equivalent choices, {\it e.g.}, another possibility is $F(u,v)=uf^{-1}(\sqrt{v^2-k^2})-v$.}
\begin{eqnarray}
F(u,v)=\sqrt{f\left(\frac{v}{u}\right)^2+k^2}-v
\end{eqnarray}
indeed satisfies Eq.(\ref{goup}).
Subsequently, the general solution of $\omega^*(k,T,\mu)$ for finite chemical potential is given by
\begin{eqnarray}
\sqrt{f\left(\frac{T\omega^*}{\omega^*-\mu}\right)^2+k^2}-\omega^*=0 ,
\end{eqnarray}
or
\begin{eqnarray}
m=f\left(\frac{T\omega^*}{\omega^*-\mu}\right) ,
\end{eqnarray}
which is Eq.(\ref{go7b}).

As for Eq.(\ref{go9}), one may immediately recognize that the equation possesses the same form of Eq.(\ref{ffchar}) by recognizing
\begin{eqnarray}
a(\mu,T,m)&=&\llangle 1 \rrangle_+ ,\nonumber \\
b(\mu,T,m)&=&-\llangle 1 \rrangle_- ,\nonumber \\
c(\mu,T,m)&=&0 .
\end{eqnarray}
Therefore, the formal solution reads
\begin{eqnarray}
\frac{d\mu}{\llangle 1 \rrangle_+}=\frac{dT}{\llangle 1 \rrangle_-} ,
\end{eqnarray}
which is the solution presented in the main text.

\bibliographystyle{h-physrev}
\bibliography{references_qian}

\end{document}